\documentclass[11pt]{article}

\usepackage{graphicx}
\graphicspath{{figures/rendered/}{figures/}}
\usepackage{bm}
\usepackage[separate-uncertainty=true, multi-part-units=single]{siunitx}[=v2]
\usepackage{amsmath, amssymb}

\renewcommand{\vec}[1]{\boldsymbol{\mathbf{#1}}} 

\title{Direct visualization of polaron formation in the thermoelectric SnSe}

\author
{Laurent P. Ren\'{e} de Cotret$^{1}$, Martin R. Otto$^{1}$, Jan-Hendrik P\"ohls$^{1,\dagger}$, \\
Zhongzhen Luo$^{2}$, Mercouri G. Kanatzidis$^{2}$, Bradley J. Siwick$^{1,3,\ast}$\\
\\
\normalsize{$^{1}$Department of Physics, Center for the Physics of Materials, McGill University, Montr\'{e}al, QC, CA}\\
\normalsize{$^{2}$Department of Chemistry, Northwestern University, Evanston, IL, USA}\\
\normalsize{$^{3}$Department of Chemistry, McGill University, Montr\'{e}al, QC, CA}\\
\normalsize{$^{\dagger}$Currently at the Department of Chemistry and Chemical Biology, McMaster University, Hamilton, ON, CA}\\
\normalsize{$^\ast$To whom correspondence should be addressed: bradley.siwick@mcgill.ca.}
}

\date{}

\begin{document}

\maketitle

\begin{abstract}
SnSe is a layered material that currently holds the record for bulk thermoelectric efficiency. The primary determinant of this high efficiency is thought to be the anomalously low thermal conductivity resulting from strong anharmonic coupling within the phonon system. Here we show that the nature of the carrier system in SnSe is also determined by strong coupling to phonons by directly visualizing polaron formation in the material.  We employ ultrafast electron diffraction and diffuse scattering to track the response of phonons in both momentum and time to the photodoping of free carriers across the bandgap, observing the bimodal and anisotropic lattice distortions that drive carrier localization. Relatively large (\SI{18.7}{\angstrom}), quasi-1D polarons are formed on the \SI{300}{\femto\second} timescale with smaller (\SI{4.2}{\angstrom}) 3D polarons taking an order of magnitude longer (\SI{4}{\pico\second}) to form.  This difference appears to be a consequence of the profoundly anisotropic electron-phonon coupling in SnSe, with strong Fr\"ohlich coupling only to zone center polar optical phonons. These results demonstrate that carriers in SnSe at optimal doping levels results in a high polaron density and that strong electron-phonon coupling is also critical to the thermoelectric performance of this benchmark material and potentially high-performance thermoelectrics more generally.
\end{abstract}

Thermoelectric materials convert a difference in temperature into an electrical potential (i.e. the thermoelectric effect) and promise to become increasingly important components of energy-harvesting devices and technologies~\cite{Sootsman2009,Snyder2008,Zebarjadi2012}. This could make a significant contribution to sustainability efforts by enabling electrical power generation from otherwise wasted heat. Unfortunately, the combination of thermal and electrical properties that lead to high thermoelectric performance is not found in naturally occurring materials~\cite{Zhu2017}; efficient thermoelectrics must be \emph{engineered}. The figure of merit for thermoelectric efficiency is $ZT = (S^2 \sigma /\kappa)T$, where $T$ is the absolute temperature, $S$ is the Seebeck coefficient (induced voltage per temperature gradient), $\sigma$ is the electrical conductivity, and $\kappa$ is the thermal conductivity. Historically, increasing $ZT$ has consisted in starting with compounds with a high Seebeck coefficients (e.g. selenides), and then selecting for good electrical conductivity and low thermal conductivity. In this sense, the ideal thermoelectric has been referred to as a ``phonon glass -- electron crystal''~\cite{Rowe1995} -- a concept that interestingly also seems to apply to lead-halide perovskite light harvesting materials~\cite{Miyata2017}. Nanostructuring has also been explored as a way to further reduce the thermal conductivity of thermoelectrics without significantly impacting electrical conductivity~\cite{Hicks1993b}, but concerns regarding manufacturing and longevity favour the use of bulk materials~\cite{Minnich2009}. In practice, attempts to optimize the performance of bulk thermoelectric materials by favourably influencing a single parameter has not been a successful strategy, because the key properties are all interdependent~\cite{Hicks1993b}. Thus, developing a more sophisticated understanding of the fundamental interdependencies between key material parameters ($S$, $\sigma$, and $\kappa$) is widely recognized as critical to the development of high-performance thermoelectrics. 

\begin{figure*}
    \centering
	\includegraphics{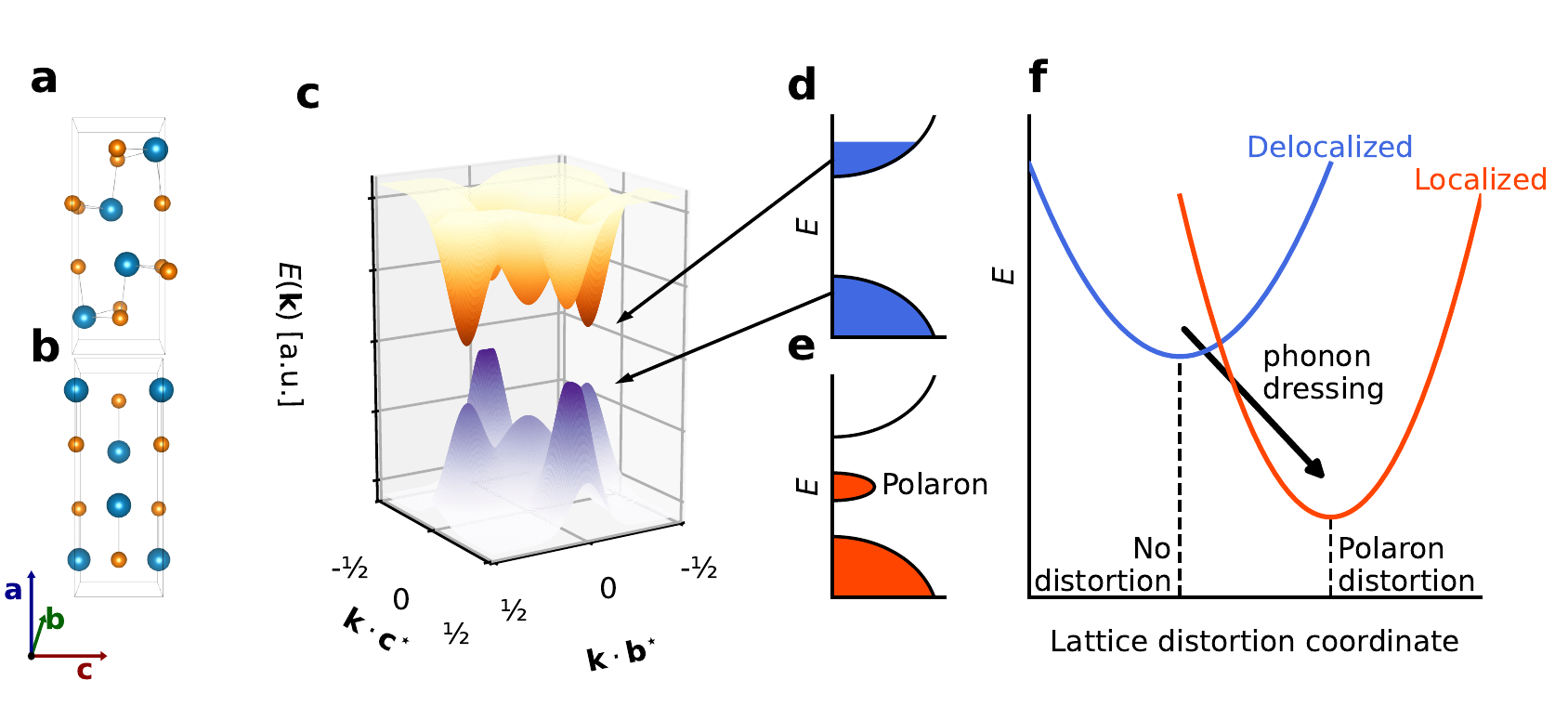}
	\caption{\textbf{a} SnSe has a layered orthorhombic structure in the $Pnma$ phase at room-temperature \textbf{b} The $Pnma$ structure is derived from 3D distortion of the higher symmetry (distorted rock-salt) $Cmcm$ phase, which is stable above \SI{750}{\kelvin}). Crystallographic directions are all given with respect to the low-temperature $Pnma$ phase and structures were rendered using VESTA~\cite{Vesta2011} \textbf{c} Diagram of the electronic structure of $Pnma$ SnSe in the $\vec{b}^\star$--$\vec{c}^\star$ plane at \SI{300}{\kelvin}. Three distinct valleys are present in both the valence and conduction bands: at $\Gamma$, $\frac{2}{3} Y$, and $\frac{3}{4} Z$~\cite{Wei2019}.  Photoexcitation at \SI{800}{\nano\meter} (\SI{1.55}{\electronvolt}) photodopes electrons and holes into all three valleys. \textbf{d} Schematic (single-valley) band structure diagram immediately following photoexcitation, which generates delocalized conduction-band electrons (equivalent picture for holes not shown). \textbf{e} Schematic band structure diagram after carrier localization indicating a polaron peak below the Fermi energy $E_F$. \textbf{f} Configuration coordinate showing the free energy of the system as carriers self-localize via phonon dressing; i.e. the generation of a local lattice distortion~\cite{Franchini2021}. The phonon wavevector dependence of this dressing process is probed directly through the UEDS experiments reported here}
	\label{FIG:structure}
\end{figure*}

Recently, tin selenide (SnSe) has been shown to exhibit remarkable thermoelectric efficiency, while also being non-lead-based and composed of earth-abundant elements~\cite{Zhao2014,Zhao2016, Zhao2016b}. Its high performance owes to three factors; i) an anomalously-low lattice thermal conductivity $<\SI{1}{\watt\per\meter\per\kelvin}$~\cite{Zhao2014} at room temperature that decreases even further at higher temperature, ii) an electrical conductivity that increases notably above \SI{600}{\kelvin} and iii) a high Seebeck coefficient. These factors combine to yield a profound enhancement in thermoelectric performance in SnSe above \SI{600}{\kelvin}, from $ZT \sim 0.1$ to a maximum $ZT > 2$ at \SI{800}{\kelvin}~\cite{Zhao2014, Zhao2016}. In the case of undoped SnSe, this enhancement is associated with a second order $Pnma \to Cmcm$ phase transition (Fig. \ref{FIG:structure} a and b) of a displacive character~\cite{Chattopadhyay1986}, but at a microscopic level is related to changes in the character of the electron-phonon~\cite{Caruso2019} and phonon-phonon~\cite{Li2015} interactions that control electrical and thermal transport in the material. 

Unlike most previous studies, which have attempted to understand the enhancement of thermoelectric properties in SnSe in terms of lattice anharmonicity and the $Pnma \to Cmcm$ phase transition~\cite{Li2015,Skelton2016,Aseginolaza2019,Lanigan2020}, in this work we focus on the momentum-dependence of electron-phonon coupling in the $Pnma$ phase specifically\cite{Li2019}. We seek to develop a full understanding of the carrier-lattice interactions that may also contribute to thermoelectric performance. To this end, we use ultrafast electron diffraction (UED) and diffuse scattering (UEDS)~\cite{Chase2016,Waldecker2017,Stern2018,RenedeCotret2019,Otto2021} to directly probe electron-phonon interactions in momentum and time following the photodoping of carriers (Fig. \ref{FIG:structure} c). A feature of these ultrafast measurements is that they freeze out changes in electronic and phonon bandstructure that results from thermal expansion (temperature-dependent lattice constants), uncoupling those effects from those due exclusively to the photodoped carriers that are the subject of our investigations.  The results presented here clearly reveal profoundly momentum-dependent electron-phonon coupling in SnSe, as is expected from Fr\"ohlich coupling in a polar lattice~\cite{Caruso2019}.  However, the ultrafast diffuse scattering signals also show clear signatures of the 'phonon dressing' (lattice distortion) that drives photo-carrier localization and polaron formation (Fig. \ref{FIG:structure} d - f), even at high levels of photo-carrier doping (equivalent to the doping levels previously used to optimize the power factor in SnSe~\cite{Zhao2016}).  
 
\subsection*{Results}

The combined UED/UEDS experiments are based on a pump-probe scheme, and were performed with a radio-frequency compressed ultrafast electron scattering instrument described in detail elsewhere~\cite{Chatelain2012, Otto2017}. Briefly, \SI{35}{\femto\second} pump pulses at \SI{800}{\nano\meter} (\SI{1.55}{\electronvolt}) drive vertical electronic transitions at $t=t_0$, photodoping electrons and holes into the band valleys near $\Gamma$, $\frac{2}{3}Y$ and $\frac{3}{4}Z$~\cite{Li2015,Zhao2016,Melendez2018,Wei2019} (Fig. \ref{FIG:structure}c). The effects of this photoexcitation on both the lattice structure and phonon system are measured with UED and UEDS simultaneously at a time $t = t_0 + \tau$. Time-series of the changes in electron scattering intensity at all scattering vectors, $\vec{q}$, were assembled by scanning the pump-probe time delay, $\tau$. The experiments were repeated over a range of pump-fluences at \SI{300}{\kelvin}.

\begin{figure*}
    \centering
	\includegraphics{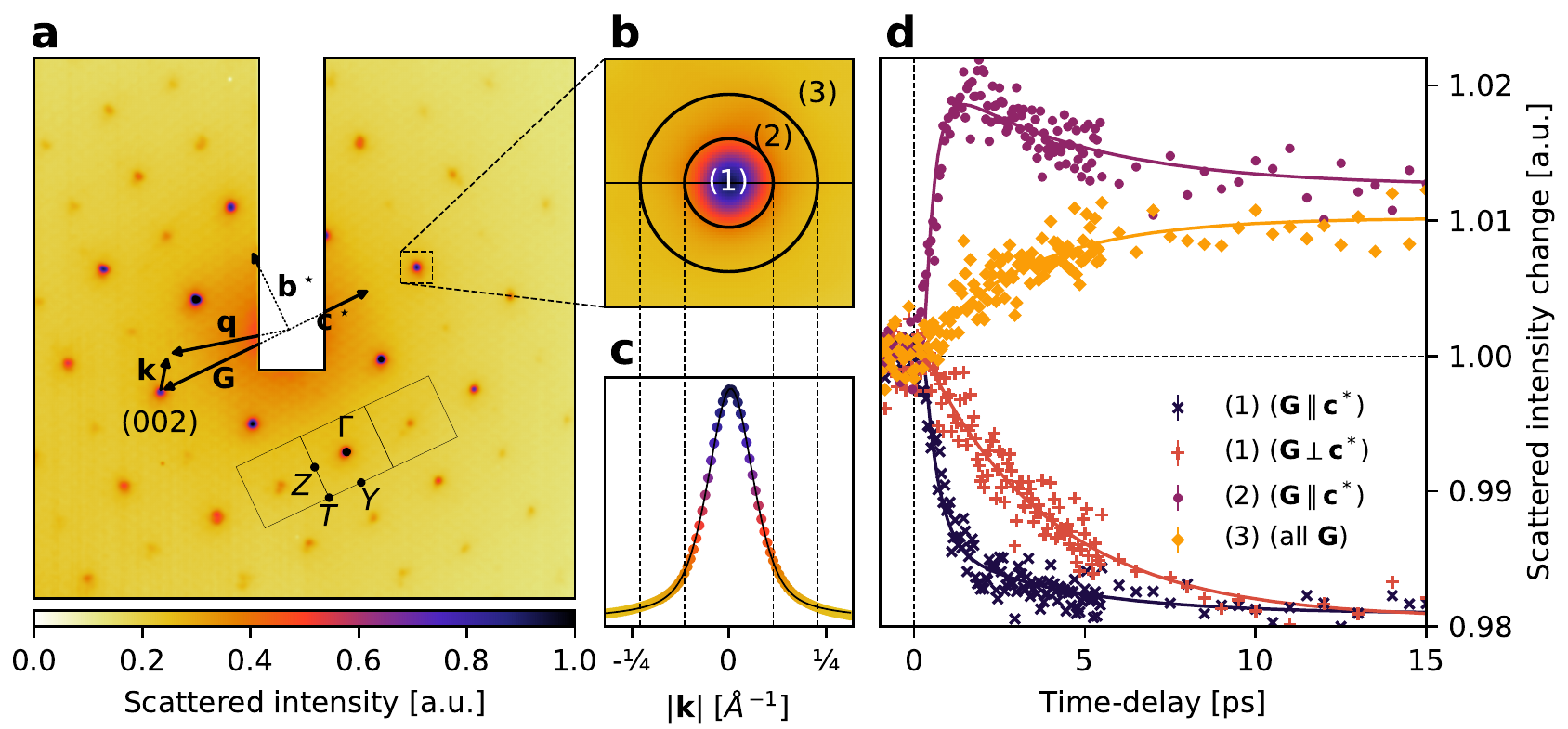}
	\caption{Ultrafast electron diffraction and diffuse scattering signals from photodoped SnSe. \textbf{a} Equilibrium scattering pattern of SnSe oriented along the $[100]$ direction with key vectors, the square BZ ($\vec{b}^\star$--$\vec{c}^\star$ plane) and high symmetry points indicated. \textbf{b} Regions of interest for scattering as described in the text, shown around reflection $(00\bar{2})$ as an example; (1) Bragg intensity, (2) small-wavevector phonons ($\SI{0.114}{\per\angstrom} <|\vec{k}| < \SI{0.228}{\per\angstrom}$) and (3) larger wavevector phonons ($|\vec{k}| > \SI{0.228}{\per\angstrom}$)  \textbf{c} Line cut across the horizontal line shown in panel b). The Bragg peak lineshape is fit with a Voigt profile (solid black line) with a full-width at half-max of \SI{0.158}{\per\angstrom}. \textbf{d} Transient (photoinduced) ultrafast electron scattering intensity changes in several regions of the BZ shown in a) and b). The decrease of intensity directly on the Bragg peaks shows transient DW decays that are strongly anisotropic. The fast ($\sim \SI{300}{\femto\second}$) decay component is maximized in Bragg peaks along $\vec{c}^\star$ (black), and only the slow components ($\sim \SI{4}{\pico\second}$) is observed in peaks perpendicular to $\vec{c}^\star$ (red). Transient diffuse intensity also shows pronounced anisotropy and $\vec{k}$-dependence. A fast rise in diffuse intensity (purple) is only observed for $|\vec{k}| < \SI{0.228}{\per\angstrom}$ (region 2, panel b) in BZs that show the fast DW dynamics. A slow increase in diffuse intensity (orange) is observed at all zone boundary high-symmetry points (see Fig. S4) and elsewhere in the BZ for all reflections $\vec{G}$ (e.g. region (3), panel b). Error bars represent the standard error in the mean of intensity before time-zero, but are generally smaller than the markers.}
	\label{FIG:data}
\end{figure*}

The total measured electron scattering intensity $I(\vec{q}, \tau)$ can be decomposed into $I(\vec{q}, \tau) = I_0(\vec{q}, \tau) + I_1(\vec{q}, \tau) + ...$, where $I_0$ corresponds to the Bragg peaks of conventional diffraction and $I_1$ is known as \emph{single-phonon diffuse scattering}. The intensity of Bragg peaks at scattering vector $\vec{G}$, $I_0(\vec{q}=\vec{G},\tau)$, reports on the lattice constants, unit cell structure, coherent modulation~\cite{Chatelain2014, Sie2019} and/or transient Debye-Waller (DW) suppression~\cite{Siwick2003, Ernstorfer2009, Waldecker2016} of peak intensity. Diffraction signals can be $10^5$--$10^8$ times more intense than phonon diffuse scattering signals~\cite{Stern2018,RenedeCotret2019}. The expression for single-phonon diffuse scattering (PDS) is given by:
\begin{equation}
	I_1(\vec{q}, \tau) \propto \sum_\lambda \left| a_{\lambda \vec{k}}\right|^2 \left| F_{1\lambda}\left(\vec{q}, \left\{ e_{\lambda\vec{k}}\right\} \right) \right|^2 \label{EQ:diffuse}
\end{equation}
where the label $\lambda$ indicates the specific phonon branch, $\vec{q}$ is the electron scattering vector, $\vec{k}$ is the reduced phonon wavevector (i.e. $\vec{k}$ = $\vec{q}$ - $\vec{G}$, where $\vec{G}$ is the closest Bragg peak), $a_{\lambda\vec{k}}$ is the vibrational amplitude of mode $\lambda$, and $F_{1\lambda}$ are known as the one-phonon structure factors. $I_1$ provides momentum-resolved information on the nonequilibrium distribution of phonons across the entire Brillouin zone, since $I_1(\vec{q}, \tau)$ depends only on phonon modes with wavevector $\vec{k}$ = $\vec{q}$ - $\vec{G}$ (Fig. \ref{FIG:data}a). The one-phonon structure factors $F_{1\lambda}$ are geometrical weights that depend sensitively on the phonon mode atomic polarization vectors $\left\{ e_{\lambda\vec{k}}\right\}$~\cite{RenedeCotret2019}. Most importantly, $F_{1\lambda}\left(\vec{q}, \left\{ e_{\lambda\vec{k}}\right\} \right)$ are relatively large when the phonon mode $\lambda$ is polarized parallel to the scattering vector $\vec{q}$. Terms of higher-order than $I_1$ have lower cross-sections and do not contribute significantly to the interpretation of the low-order Brillouin zone (BZ) signals reported on here~\cite{Wang2013,Zacharias2021a,Zacharias2021b}. 

\begin{figure*}
    \centering
    \includegraphics{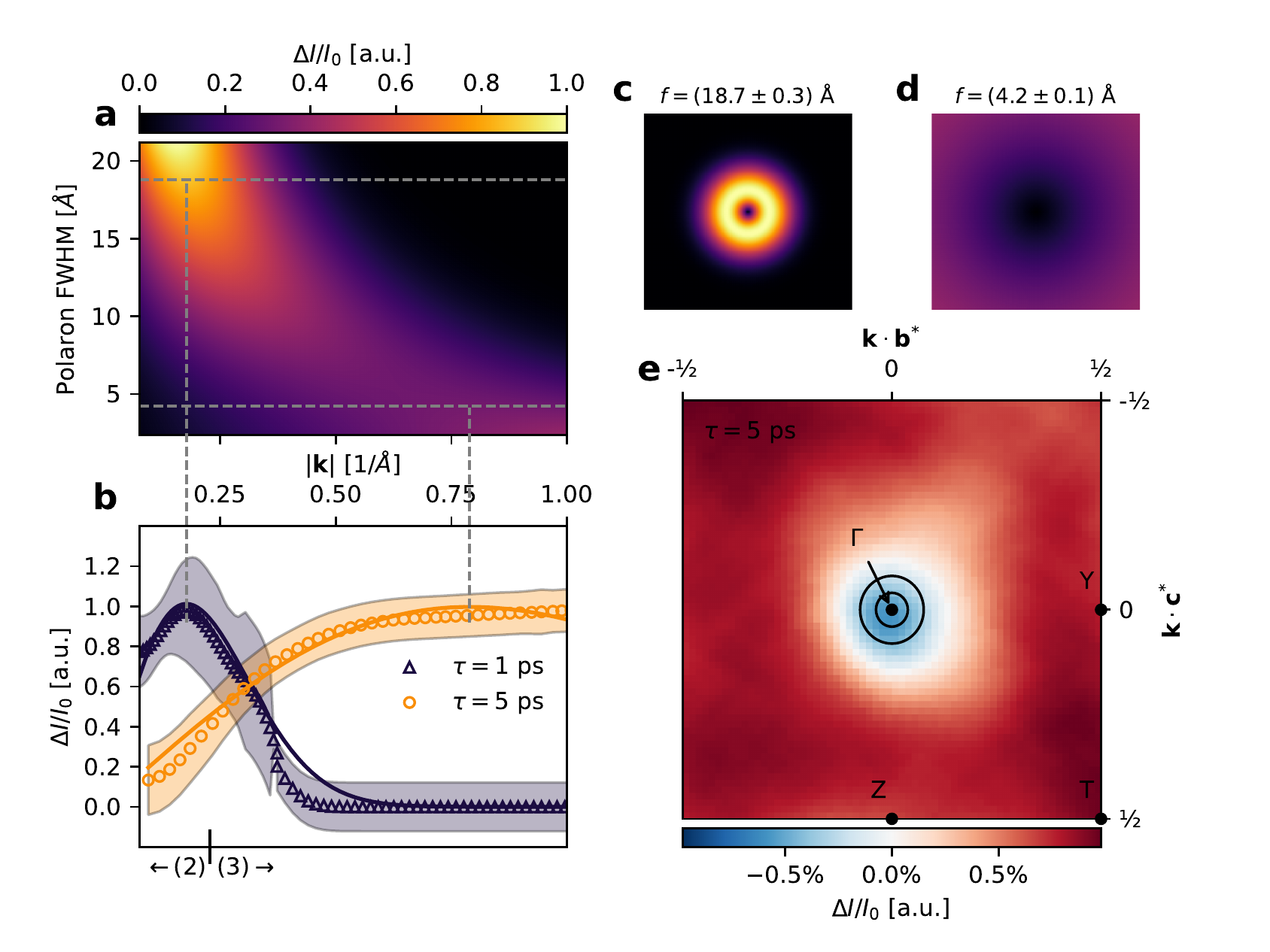}
    \caption{Polaron formation in SnSe visualized with UEDS. \textbf{a} Wavevector-dependent scattering intensity for a Gaussian displacement field model of the polaron lattice distortion as a function of size (polaron FWHM), as described in the text. \textbf{b} Measured change in diffuse intensity at \SI{1}{\pico\second} (black triangles) and \SI{5}{\pico\second} (orange circles) fit to the Gaussian displacement model above (solid curves). Best-fit FWHM polaron dimensions are $f=\SI{18.7 \pm 0.3}{\angstrom}$ (\SI{1}{\pico\second}) and $f=\SI{4.2 \pm 0.1}{\angstrom}$ (\SI{5}{\pico\second}).  The larger polaron is consistent with uniaxial displacements along $\vec{c}^\star$ and the smaller polaron with uniform displacements in the $\vec{b}$--$\vec{c}$ plane. The boundary between regions (2) and (3) from Fig. \ref{FIG:data}b are indicated. The shaded region represents the standard error in the mean of intensity across the integration region. \textbf{c} Differential scattering intensity across the BZ due to the large polaron lattice distortion. \textbf{d} Differential scattering intensity across the BZ due to the small polaron lattice distortion. \textbf{e} The measured differential diffuse intensity across the BZ at $\tau=\SI{5}{\pico\second}$ is in excellent agreement with the predicted scattering vector dependence of the model.}
    \label{FIG:polaron}
\end{figure*}

Fig. \ref{FIG:data}a shows an equilibrium diffraction pattern of SnSe at room temperature along the $[100]$ zone axis with the rectangular in-plane ($\vec{b}^\star$--$\vec{c}^\star$) BZ of the $Pnma$ phase indicated. This pattern contains both Bragg scattering (at zone-center positions) and PDS contributions at all scattering vectors.  Following photoexcitation, Bragg peaks show transient decreases in intensity whose dynamics are well-described by a biexponential decay with time-constants (\SI{400 \pm 130}{\femto\second} and \SI{4 \pm 1}{\pico\second}), as shown in Fig. \ref{FIG:data}d.  The fast component of these dynamics are profoundly anisotropic, with a maximum contribution in the $\vec{c}^\star$ direction and below detection along $\vec{b}^\star$ (Fig. \ref{FIG:data}d).  Given the DW factor ($\exp(-\frac{1}{2}\langle \vec{q} \cdot \vec{u} \rangle^2$), this indicates that there are at least two distinct processes that contribute to increasing the atomic displacement $\vec{u}$ following photoexcitation.  The ultrafast dynamics of the PDS intensity following photoexcitation provides a clear perspective on these distinct processes. At all high-symmetry BZ boundary positions ($Z$, $Y$ and $T$) we find that PDS intensity increases with a single exponential time constant of \SI{4 \pm 1}{\pico\second} (Fig. S4) that is the complement of the slow time-constant observed in the Bragg peak dynamics. In fact, within experimental uncertainties an identical time-constant is determined for increases in PDS observed at all BZ positions $|\vec{k}| > \SI{0.228}{\per\angstrom}$ (i.e. far from Bragg peaks), as is shown in Fig. \ref{FIG:data}d (yellow) and Fig. \ref{FIG:polaron}e. 

Previous work has identified a number of soft and strongly-coupled optical phonons in the zone-center region of the $Pnma$ phase~\cite{Chattopadhyay1986,Li2015,Gong2020,Lanigan2020}.  As $\vec{k}$ approaches zone-center, the PDS contribution overlaps with the Bragg peak lineshape.  Thus, Bragg peak intensity must be subtracted to accurately determine the differential PDS from small-wavevector phonons following photoexcitation. As part of this analysis we investigated whether photoexcitation resulted in measurable time-dependence of Bragg peak positions and widths.  Fig. S3 demonstrates that the Bragg peak center positions and widths are shot noise limited and neither parameter shows a measurable time-dependence;  i.e. any photoinduced change to in-plane lattice constants (which shift Bragg peak positions) or in-plane long-range strain (which broaden and skew the width of peaks) over the range of delay times investigated here are at a level that is below our signal-to-noise ratio. The Bragg peak lineshapes are effectively constant, but have integrated intensities that vary according to the observed transient Debye-Waller dynamics. Panels b and c of Fig. \ref{FIG:data} show two regions of interest around every Bragg peak. We define an area at the BZ center associated with strongest Bragg scattering (region (1), $|\vec{k}| \leq \SI{0.114}{\per\angstrom}$) and a surrounding region associated with wavevectors in the range $\SI{0.114}{\per\angstrom} < |\vec{k}| \leq \SI{0.228}{\per\angstrom}$ (region (2)). By integrating scattered intensity in these areas separately, we assemble time-series that probe both the Bragg (region 1) and the small wavevector ($|\vec{k}| \sim \Gamma$) PDS (region 2) after subtraction of Bragg intensity.  The results for reflections parallel to $\vec{c}^\star$ is shown in Fig. \ref{FIG:data}d (purple), demonstrating that the fast component of the DW dynamics is exclusively associated with a rapid increase in the amplitude of phonons near zone-center. This signal shows the same $\vec{b}^\star$/$\vec{c}^\star$ anisotropy as the Bragg peaks (Fig. S6) and is \emph{insensitive} to the precise definition of the indicated regions, as demonstrated below.  Determination of the relatively small differential PDS signals directly under the Bragg peak (region 1) is obviously subject to large uncertainties and is not reported here. These observations corroborate computational work~\cite{Caruso2019,Ma2018} which found that the electron-phonon coupling is profoundly anisotropic in SnSe, with carriers coupling very strongly to polar zone-center modes and only much more weakly elsewhere. 

The PDS in intermediate regions of the BZ is a mixture of the fast and slow dynamics shown in Fig. \ref{FIG:data}d, with the magnitude of these components varying as a function of $|\vec{k}|$ as shown in Fig. \ref{FIG:polaron}. In a conventional weakly polar semiconductor like GaAs, these dynamics can be described in terms of a model for the re-equilibration of the photodoped carriers with the phonon-system via intra- and inter-valley inelastic electron-phonon scattering processes based on the electron and phonon bandstructures of the material~\cite{Sjakste2018}.  The dynamics observed here for SnSe cannot be described in these terms as is explained further in the discussion below.  Polaron formation, however, provides a robust description of these observations.      

\begin{figure}
    \centering
    \includegraphics{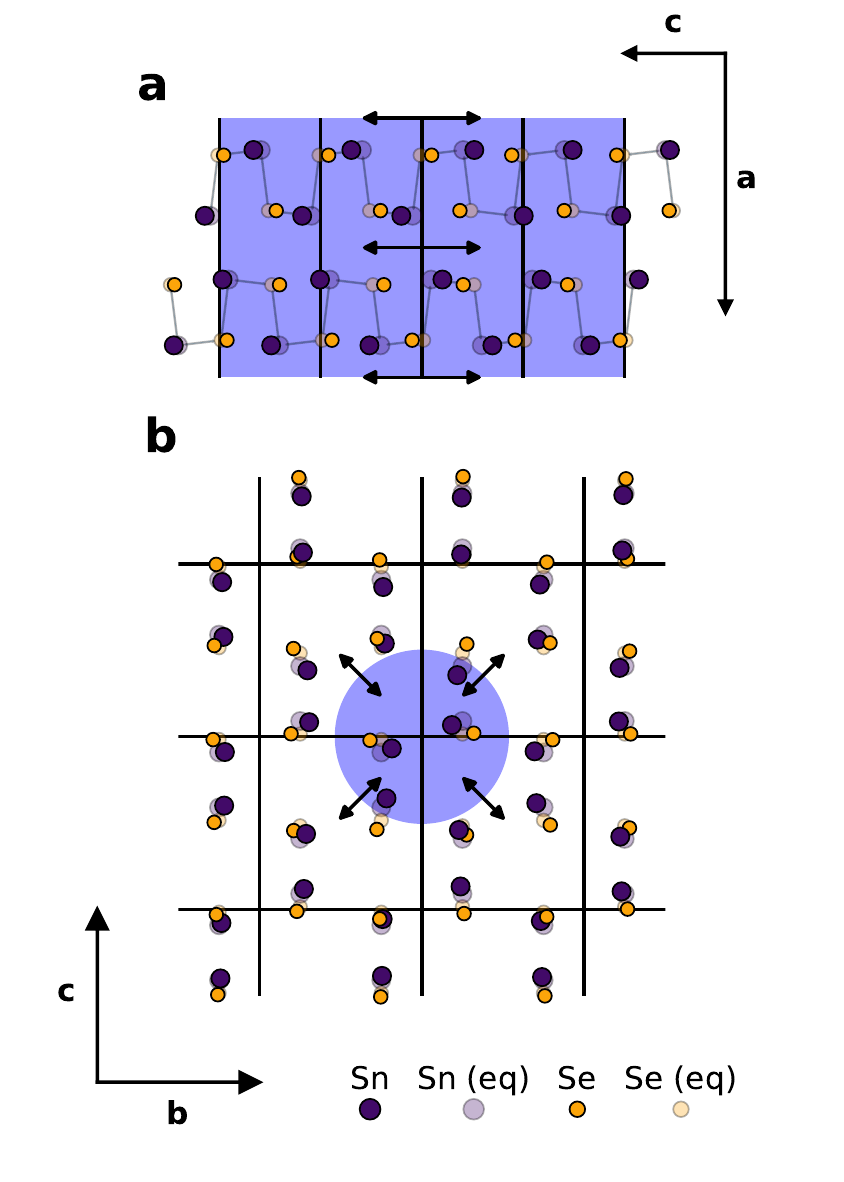}
    \caption{Real-space visualization of the in-plane ($\vec{b}$ -- $\vec{c}$) atomic displacement due to the large and small polarons in Fig. \ref{FIG:polaron}. The unperturbed in-plane dimensions of the unit cell are marked by solid lines. \textbf{a} Large (\SI{18.7}{\angstrom} FWHM) one-dimensional polaron aligned to the $c$-axis \textbf{b} Small (\SI{4.2}{\angstrom} FWHM) three-dimensional polaron. In both subpanels, the blue-shaded background represents the FWHM of the atomic displacement field $\vec{u}(\vec{r})$. The magnitude of the atomic displacements has been exaggerated for visual clarity.}
    \label{FIG:polaron-realspace}
\end{figure}

The local lattice distortions associated with a polaron in real-space can be modelled as an atomic displacement field $\vec{u}(\vec{r})$, where atoms are displaced as $\vec{r}_j \to \vec{r}_j + \vec{u}(\vec{r}_j)$ \cite{Guzelturk2021}. For small displacement fields, the contribution of such a localized lattice distortion to the total scattering amplitude, $f^p$, is given by:
\begin{equation}
    f^p(\vec{q}) \approx - i \sum_j f_{e,j}(\vec{q}) e^{-i \vec{q} \cdot \vec{r}_j} \left( \vec{G} \cdot \vec{u}_j \right)
\end{equation}
where $\left\{ \vec{r}_j \right\}$ are the atomic positions, $\left\{ f_{e,j} \right\}$ are the atomic form factors for electron scattering, and $\vec{G}$ is the Bragg reflection nearest to $\vec{q}$ as in Fig. \ref{FIG:data} (see SI). We consider two specific displacement fields due to the anisotropic observations described above; an effectively one-dimensional displacement field directed along the $c$-axis ($\vec{u}_c(\vec{r}) \propto e^{-|\vec{r}|^2/r_p^2} ~ \hat{\vec{r}} \cdot \hat{\vec{c}}$), and a three-dimensional displacement field ($\vec{u}(\vec{r}) \propto e^{-|\vec{r}|^2/r_p^2} \hat{\vec{r}}$) (see SI). In both cases, $2 \sqrt{2 \ln(2)} ~ r_p$ is the full-width at half-maximum (FWHM) of the local lattice distortion associated with the polaron. The effect of both displacement fields is identical in terms of the impact on the diffuse scattering measured within a BZ, shown as a function of polaron FWHM in Fig. \ref{FIG:polaron} a).  The one-dimensional displacement field, however, has the same $\vec{b}^\star$/$\vec{c}^\star$ anisotropy in scattering intensity as our measurements.

This model was fit to the measured photoinduced differential PDS signals at \SI{1}{\pico\second} and at \SI{5}{\pico\second}, to capture changes due to the fast and slow dynamics respectively. The fast PDS dynamics are in excellent agreement with the formation of a \SI{18.7 \pm 0.3}{\angstrom} polaron displacement field. The slow PDS dynamics, by contrast, are in excellent agreement with the displacement field associated with a \SI{4.2 \pm 0.1}{\angstrom} FWHM polaron. A qualitative real-space representation of the two polaron modes is presented in Fig. \ref{FIG:polaron-realspace}. We tentatively assign the larger polaron to the electron and the smaller polaron to the hole, in analogy to the work on Sio \emph{et al.}~\cite{Sio2019} on other polar materials. However, these observations are also consistent with electron and hole polarons being a similar size in SnSe (i.e. \SI{4.2 \pm 0.1}{\angstrom} FWHM) and formation occurring in two steps.  We discuss these results within the context of both interpretations below.  

\subsection*{Discussion}

 The carriers generated through photoexcitation in these experiments localize via their interactions with the phonon system.  These interactions create a local potential that minimizes the free energy of the system according to the standard picture of polaron formation ~\cite{Franchini2021} shown schematically in Fig. 1 d) - f). A polaron quasiparticle is thus formed as a localized carrier dressed by phonons, potentially in several phonon branches and over a range of wavevectors~\cite{Sio2019}; a local lattice distortion in real-space is equivalent to a distribution of lattice normal modes in reciprocal space. These details are revealed by the UED and UEDS data. As recent ab initio work by Giustino and colleagues has demonstrated in other polar lattices~\cite{Sio2019}, polarons require the recruitment of phonon modes across the entire BZ when the dimensions of the polaron approaches those of the lattice constants.  In SnSe, we propose that this manifests in the bimodal formation dynamics reported here due to the profoundly anisotropic (momentum-dependent) nature of electron-phonon coupling in the material.  Strong Fr\"ohlich coupling to near zone-center polar optical phonons ($A_g$, $B_u$, and $B_g$) results in the rapid ($\sim\SI{300}{\femto\second}$) formation of relatively large (electron) polarons, since polarons of this size only require the recruitment of strongly coupled small-wavevector phonon modes.  The formation time of the smaller (hole) polarons is an order of magnitude longer ($\sim \SI{4}{\pico\second}$) due to the relatively weak coupling to the large-wavevector phonons that must be recruited to form polarons of this size~\cite{Sio2019}. 

Alternatively, the observed bimodal polaron formation dynamics are also consistent with a two-step process that has features in common with Onsager's inverse snowball effect, often discussed in the context of the theory of solvation~\cite{neria1992simulations}.  Here the rapid but weak localization of the charge carriers is provided by the 1D ferroelectric-like lattice distortions along the $c$-axis.  The slower but strong localization is provided by the subsequent 3D distortions.  Polaron formation dynamically proceeds from the outside (long range) inwards (short range), not via layer accumulation from the inside out like a typical snowball.    

\begin{figure}
    \centering
    \includegraphics{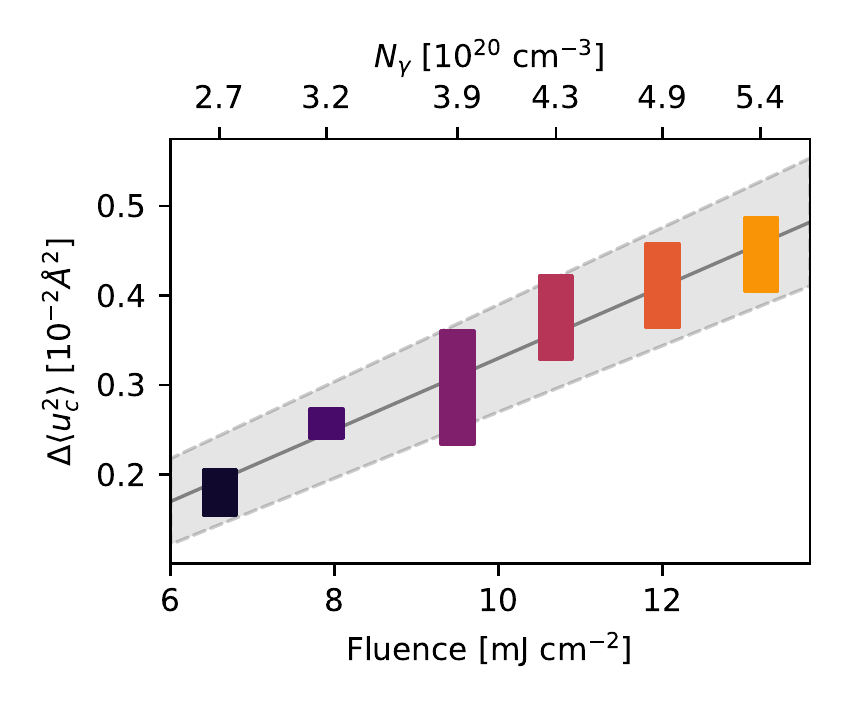}
    \caption{Increase in mean-square-displacement of all atoms along the $c$ axis, $\Delta \langle u_c^2 \rangle$, due to the change in vibrational amplitude of the strongly-coupled zone-center modes exclusively. The associated photocarrier concentration $N_{\gamma}$ for a sample of dimensions $\SI{50 x 50 x 0.045}{\micro\meter}$ is shown above the plot. Boxes are used to represent error bars along both axes.}
    \label{FIG:msd}
\end{figure}

These measurements alone cannot identify precisely how vibrational excitation is distributed over the phonon branches near zone-center by $\sim \SI{1}{\pico\second}$; however, the increase in mean-square atomic displacements can be determined from the Bragg peak DW decays and can be used to estimate the fraction of excitation energy that has left the carrier system in the form of the polaronic lattice distortion by $\sim \SI{1}{\pico\second}$. This quantity is shown in Fig. \ref{FIG:msd}, and is linear with pump fluence over the range investigated.  The observed increase, $\Delta \langle u_c^2\rangle$, is consistent with a nearly complete transfer ($>85\%$) of the excess photodoped carrier energy to zone center phonons (see SI and Fig. S7). Thus, the coupled electron-phonon system at $\sim$\SI{1}{\pico\second} is well described as a state in which the photodoped carriers in each valley have reached an equilibrium with only this limited set of strongly coupled small-wavevector phonons.  This equilibrium is well described by the formation of polaron quasiparticles (Fig. \ref{FIG:polaron} b) rather than the simple heating of the phonon system as has been observed in other systems, like graphite where rapid electron cooling through a pre-equilibrium with specific strongly coupled modes has been observed ~\cite{Stern2018,RenedeCotret2019}.  A similarly rapid coupling of electronic excitation energy to polaronic lattice distortions is thought to be present in methylammonium lead iodide perovskites~\cite{Niesner2016}.

Polaron formation is the simplest \emph{self-consistent} explanation of these data. The isotropic slow-rise in diffuse scattering observed across the entire BZ specifically precludes an understanding of these results in terms of the conventional semiconductor picture of carrier relaxation through intervalley scattering mediated by large-wavevector phonons~\cite{Sjakste2018,Waldecker2017,Stern2018,RenedeCotret2019,Otto2021}.  Based on the electronic dispersion of SnSe calculated by Wei \emph{et al.}~\cite{Wei2019}, we modelled the decay of hot electrons and holes, mediated by phonons, via energy-allowed and momentum-conserving pathways. This relaxation mechanism imprints the structure of the electronic dispersion onto the PDS, including a pronounced anisotropy between the $\vec{b}^\star$ and $\vec{c}^\star$ directions as shown in Fig. S8. This is ruled out by our measurements, which are azimuthally symmetric in reciprocal space (Fig. \ref{FIG:polaron}e). Neither does the anharmonic decay of phonons provide an explanation for these data, given that phonon lifetimes are estimated to be almost an order of magnitude longer (\SIrange{15}{30}{\pico\second})~\cite{Chandrasekhar1977,Li2015,Lanigan2020} than the timescales observed herein. Moreover, the anharmonic decay of phonons measured in the time-domain displays an imprint of the phonon dispersion due to energy- and momentum-conservation rules~\cite{Stern2018,RenedeCotret2019}, which is not seen in our data (Fig. \ref{FIG:polaron}e).

There are a number of important connections between these observations and an understanding of the thermoelectric properties of SnSe.  First, the rate of large polaron formation is very rapid and appears to be at or near the limit imposed by the period of the highest frequency phonons in SnSe ($\sim \SI{200}{\femto\second}$) due to the strong Fr\"ohlich coupling in the polar lattice, even at the high carrier density operating conditions of optimally doped SnSe for thermoelectric device applications~\cite{Sootsman2009,Zhao2016,Fan2018}.  At this doping level, these results indicate that SnSe is well described as a dense polaron system, since the density of polarons overlapping at a distance equal to their FWHM is  \SI{3.2e21}{\per\cubic\centi\meter} for the smaller (hole) polarons, overlapping with this optimal doping range.  
The polaronic nature of the charge carriers in SnSe likely plays an important role in preserving the electron-crystal, phonon-glass conditions that are important for high-performance thermoelectric materials~\cite{Rowe1995}. The dressed charges are better screened from scattering mechanisms that could otherwise deteriorate mobility at high carrier density and temperatures approaching a structural phase transition, where $ZT$ in SnSe is highest.  Several open questions remain regarding the nature of polarons in SnSe. Our measurements directly probe the in-plane lattice distortions, but do not provide information on the inter-layer (or $a$-axis) dependence. There is also the important question regarding the potential ferroelectric nature of polarons in SnSe, as has been discussed in the context of lead halide perovskites~\cite{Frost2017,Joshi2019}.  Monolayer SnSe has been investigated as a possible platform for ultrathin ferroelectrics~\cite{Fei2016}, and polar nanodomain formation and interlayer coupling may contribute to the dynamics we have observed. The in-plane polarization within a single layer can be modulated simply through changes in the angle of the Sn-Se bonds relative to the $a$-axis ~\cite{Fei2016} in a manner that is consistent with the local polaronic displacements shown schematically in Fig. \ref{FIG:polaron-realspace}.

Electron-phonon coupling is not normally considered to be an important contributor to lattice thermal conductivity, however, previous work has shown that these interactions can play a role in suppressing thermal conductivity and enhancing thermoelectric performance in Si~\cite{Liao2015} and SiGe~\cite{Fan2018} at high carrier doping.  While a giant lattice anharmonicity and 3-phonon scattering processes seems to be sufficient to explain the ultralow thermal conductivity of undoped SnSe~\cite{Zhao2014, Li2015}, a quantitative understanding of electron-phonon interactions and their impact on both electronic and lattice thermal conductivity (in addition to electrical conductivity) in this strongly coupled, high-carrier density regime is important for the further development of thermoelectrics and higher $ZT$.  The rate of electron-phonon scattering with the strongly coupled polar modes is more than an order of magnitude higher than 3-phonon scattering processes at the carrier densities investigated here~\cite{Lanigan2020}.

We are currently in an excellent position to make significant progress understanding this complex, strongly-coupled regime across many material classes. The recent parallel development of ab initio approaches for both electron-phonon coupling~\cite{Ponce2016} and polaron formation~\cite{Sio2019} and time- and momentum-resolved measurement techniques like ultrafast electron/xray diffuse scattering that are capable of interrogating electron-phonon interactions in exquisite detail and directly visualizing polaron formation.  Future work that combines these approaches together with more well established ultrafast spectroscopic methods are likely to yield significant insights.

\subsection*{Synthesis and sample preparation} 

A SnSe ingot (\SI{20}{\gram}) was synthesized by mixing appropriate ratios of high purity starting materials (Sn chunk, 99.999\%, American Elements, USA and Se shot, 99.999\%, 5N Plus, Canada) in \SI{13}{\milli\meter} diameter quartz tube. The tube was flame-sealed at a residual pressure of $\SI{1e-4}{\mmHg}$, then slowly heated to \SI{1223}{\kelvin} over \SI{10}{\hour}, soaked at this temperature for \SI{6}{\hour} and subsequently furnace cooled to room temperature. The obtained ingot was crushed into powder and flame-sealed in a quartz tube, which was placed into another, bigger, flame-sealed quartz tube. A crystal with dimensions of $\sim$\SI{13}{\milli\meter} (diameter) $\times$ \SI{20}{\milli\meter} (length) was obtained.

Seven samples were used for the ultrafast electron scattering measurements, to ensure reproducibility. Six samples were ultramicrotomed with a \ang{35} diamond blade, while one sample was mechanically exfoliated. Three of the ultramicrotomed samples were cut from a first SnSe mother flake at a thickness of \SI{90}{\nano\meter}. The remaining three ultramicrotomed samples were cut to a thickness of \SI{70}{\nano\meter} from a different mother flake, synthesized separately from the first. The exfoliated sample was prepared from the second mother crystal, with a final thickness of \SI{45}{\nano\meter} as estimated based on the relative ratio of intensities, compared to thicker ultramicrotomed samples.

\subsection*{Ultrafast electron scattering experiments} 
Electron scattering experiments employed \SI{35}{\femto\second} pulses of \SI{800}{\nano\meter} light at a repetition rate of \SI{1}{\kilo\hertz}. Part of these light pulses is upconverted to \SI{266}{\nano\meter} pulses via third-harmonic generation, which is then used to generate bunches of $10^6$ -- $10^7$ electrons from a bulk copper photocathode. These electrons are then accelerated to \SI{90}{\kilo\electronvolt} in a DC electric field. Electron bunches are compressed with a radio-frequency cavity to counterbalance space-charge repulsion, resulting in a time-resolution of $\SI{130}{\femto\second}$. Electrons are transmitted through the samples before being collected by an electron camera. The other part of the \SI{800}{\nano\meter} pulses is used to photoexcite the sample almost co-linearly with the electron propagation axis ($\sim \ang{5}$). Experiments were repeated for up to \SI{72}{\hour} to maximize signal-to-noise, which was made possible by improvements in RF-laser synchronization~\cite{Otto2017}. Detailed descriptions of this instrument are presented elsewhere~\cite{Chatelain2012, Otto2017}. The samples were thinner than than the optical depth of $>\SI{100}{\nano\meter}$ at \SI{800}{\nano\meter}~\cite{Barrios2014, Makinistian2009}, ensuring that the entire sample was photoexcited. The samples were oriented in the $\langle 100 \rangle$ direction, giving UEDS measurements a full view of lattice dynamics in the plane spanned by $\vec{b}$ and $\vec{c}$.

\bibliography{biblio}
\bibliographystyle{plain}

\end{document}